\documentclass[twocolumn,prb,nobibnotes]{revtex4}
\usepackage{revsymb}
\usepackage{graphicx} 
\usepackage{amsmath}
  \usepackage{tikz}

\usepackage{setspace}

\DeclareGraphicsExtensions{.jpg,.png}
\usepackage[pdftex,colorlinks=true,linkcolor=blue,citecolor=blue,urlcolor=blue]{hyperref}

\newcommand{\beq}{\begin{equation}}
\newcommand{\eeq}{\end{equation}}

 \newcommand{\bfi}{\begin{figure}[h]}
 \newcommand{\efi}{\end{figure}}

\begin{document}


\title{Critical Stokes number for the  capture of inertial particles by recirculation cells in 2D quasi-steady flows}

\author{Romuald Verjus}
\address{Laboratoire Protec'Som, Valognes, France}

\author{Jean-R\'{e}gis Angilella\footnote{Corresponding author, Jean-Regis.Angilella@unicaen.fr}}
\address{LUSAC-ESIX,  Cherbourg, Universit\'e de Caen Normandie, France}

\vskip.5cm
\begin{abstract}
{ Inertial particles are often observed to be trapped, temporarily or permanently, by recirculation cells which are ubiquitous in natural or industrial flows.  In the limit of small particle inertia, determining the conditions of trapping is a challenging task, as it requires a large number of numerical simulations or experiments to test various particle sizes or densities. Here, we investigate this phenomenon analytically and numerically in the case of heavy particles (e.g. aerosols) at low Reynolds number, to derive a trapping criterion that can be used both in analytical and numerical  velocity fields. The resulting criterion allows to predict the characteristics of trapped particles as soon as single-phase simulations of the flow are performed. Our analysis is valid for two-dimensional particle-laden flows in the vertical plane, in the limit where the particle inertia,  the free-fall terminal velocity, and the flow unsteadiness  can be treated as perturbations.   The weak unsteadiness of the flow generally induces a  chaotic tangle  near  heteroclinic or homoclinic cycles if any, leading to the apparent diffusion of fluid elements through the boundary of the cell.  The critical particle Stokes number $St_c$ below which aerosols also enter and exit the cell in a complex manner has been derived analytically, in terms of the flow characteristics. It involves the non-dimensional curvature-weighted  integral of the squared velocity of the steady fluid flow along the dividing streamline of the recirculation cell.     When  the flow is unsteady and $St > St_c$, a regular motion takes place due to gravity and centrifugal effects, like in the steady case. Particles driven towards the interior of the cell are trapped permanently.  In contrast, when the flow is unsteady and $St < St_c$, particles wander in a chaotic manner in the vicinity of the border of the cell, and can escape the cell.
 
}
\end{abstract}
 
\maketitle

  




  
 \section{Introduction}
   
  The settling of inertial particles in  plane two-dimensional flows characterized by quasi-steady  recirculation cells is  a situation of wide interest in natural or industrial flows. These cells can be due for example to the presence of obstacles of any kind, to temperature gradients, density gradients, or wind on water flows (see for example  Stommel  \cite{Stommel1949}, Chen \& Fung  \cite{Chan1999}, Fung  \cite{Fung1997}, Maxey \& Corrsin  \cite{Maxey1986}, Cerisier {\it et al.}  \cite{Cerisier2005}, to cite but a few). They are characterized by well-defined scales and shapes, and their lifetime is much larger than their turnover time. They have been shown to play a key role in the dynamics of inertial particles, that is tiny objects which do not follow exactly the fluid motion, like aerosols, sediments or even biological objects like plankton  \cite{Stommel1949}. Particles which penetrate into these cells  often perform a few rotations, then exit or get deposited on the walls if any. They can enter under the effect of gravity, inertia, or under the effect of any slight perturbation occurring while the particle passes near the boundary of the recirculation cell. This trapping process can be permanent, for example if particles get deposited on  walls, or temporary if particles exit the cell.
  
  In most situations, theoretical issues related to the complex dynamics of inertial particles can be investigated with a reasonable accuracy by means of numerical computations or analytical models (see for example the review paper by Cartwright {\it et al.} \cite{Cartwright2010}). The former generally requires heavy computing resources, and consists in solving the Navier-Stokes equation (for the fluid motion) coupled to Lagrangian or Eulerian models for the dispersed phase. In most cases, such computations are time-consuming, even if particles do not affect the flow. The prediction of the degree of contamination of recirculation cells by inertial particles, which requires numerous computations with varying parameters, is therefore a challenging task. In this paper, we combine analytical and numerical methods to  derive an analytical expression for the critical Stokes number (and therefore the critical size or mass) of particles which are likely to contaminate a given cell. This method is valid if the flow field can be written as a steady mean flow plus any small time-dependent perturbations.  
  
  The work is presented in three steps. First, we consider the transport equations and derive an analytical criterion allowing to determine the conditions under which inertial particles cross the separatrix, that is the limiting streamline between the cell and the rest of the flow domain (section \ref{Theory}). It is then used to predict particle capture in a simple analytical flow (section \ref{illustrationAnalyt}). Finally, the criterion is applied to accurate numerical solutions of the Navier-Stokes equations, coupled to a Lagrangian particle tracking algorithm, in section \ref{Numerical}.

\section{Asymptotic analysis of separatrix crossing}
  \label{Theory}
  
The structure of steady plane incompressible flows is the classical structure of 2D volume preserving dynamical systems, where the hamiltonian corresponds to the streamfunction  \cite{Ottino1989}.  These flows are therefore composed of saddle or elliptic stagnation points, the former being joined by streamlines, and the latter being surrounded by them.
 The recirculation cells are the elliptic stagnation points and the closed streamlines around them. The "boundaries" of the cells are formed by limiting streamlines  joining stagnation points,
 and forming homoclinic or heteroclinic cycles. These cycles are separatrices  playing a key role in the momentary or permanent trapping  of inertial particles.

\begin{figure}
\hspace*{-3cm}   {\includegraphics[width = 0.5\textwidth]{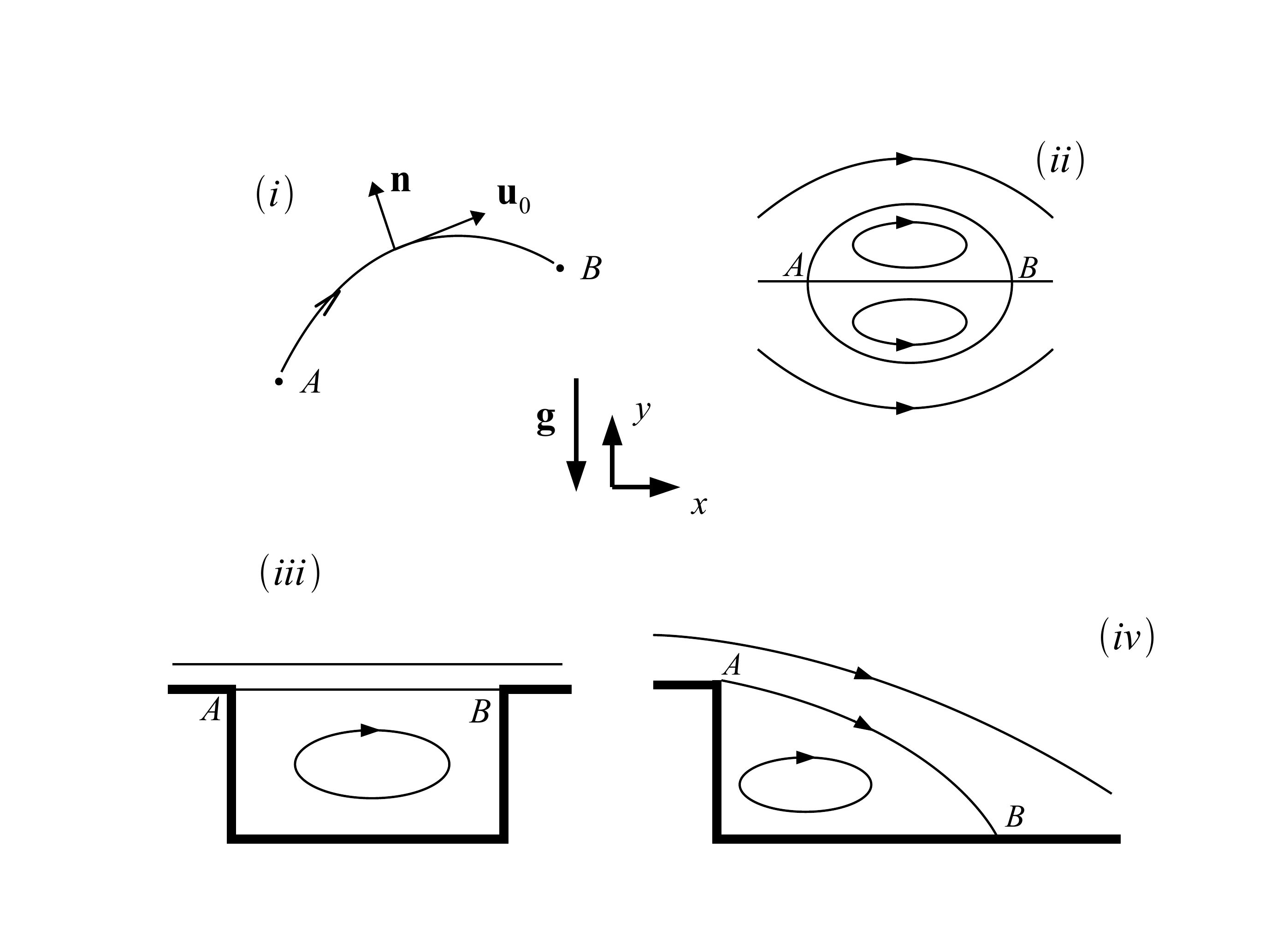}} 
\caption{\it Sketch of the stagnation points $A$ and $B$ of the steady flow $\mathbf u_0$, joined by a separatrix streamline (panel (i)). Vector $\mathbf n$ is the unit vector perpendicular to $\mathbf u_0$, such that ($\mathbf u_0,\mathbf n$) is anti-clockwise about the $z$ axis.  Some examples which will be treated in this paper: circular cell (ii), cavity flow  (iii),   backward facing-step (iv).  } 
\label{Dessin}
\end{figure}
 
Let  $L$ be the typical length scale of the flow (e.g. the overall scale of a recirculation cell), and let $U$ be the typical velocity of the flow. The non-dimensional motion equation of non-Brownian spherical particles with a small Reynolds number and a density $\rho_p$ much larger than the fluid density $\rho_f$ reads
\beq
\frac{d \mathbf v}{dt} =\frac{1}{St}(\mathbf u -\mathbf v)+ \frac{1}{Fr} \, \hat {\mathbf g} \,  
\label{EqAerosol}
\eeq  
where $\mathbf v$ is the velocity of the center-of-mass of the particle, $St= D^2 \rho_p U /(18 \mu L)$ is the Stokes number (where $\mu$ is the dynamic viscosity of the fluid), $Fr = U^2/(g L)$ is the Froude number, and $\hat {\mathbf g}$ is the unit vector in
the direction of gravity. The generalization of this theory to  particles with a density of the order of that of the fluid, like sediments in water, is among the perspectives of this work. 
Also, particle/wall interactions, either by direct contact, electrostatic forces or hydrodynamic interactions, will be neglected. We will assume that particles can slip on the walls, if any. 
Throughout the paper we will consider that particles have a finite but small inertia, that is
$$
0 < St \ll 1.
$$
The flow is composed of a steady component $\mathbf u_0$ plus a small unsteady component $\epsilon\mathbf u_1$ (with $0<\epsilon \ll 1$).
This unsteady term can be due to some noise, or to ambient turbulence with a weak turbulence intensity.
\noindent
We exploit the fact that $St \ll 1$ and look for a solution of Eq.\ (\ref{EqAerosol}) of the form 
\beq
 \mathbf v = \mathbf u_0 + St \, \mathbf v_1 + O(St^2).
 \label{expansion}
\eeq 
In addition, we assume that even though unsteadiness, inertia and sedimentation are small perturbations, none of these effects dominates the other two. Therefore, $St$ is of the order of the amplitude of the unsteady component, that is 
$St = O(\epsilon)$,
and the free-fall velocity $St/Fr$ is also of the order of the unsteady component $\epsilon\mathbf u_1$. In particular, the Froude number $Fr$ is held fixed as $\epsilon$ and  $St$ tend to zero.
Approximation (\ref{expansion}) manifests the fact that if particles are sufficiently small, their velocity $\mathbf{v}(t)$ is always close to the steady fluid velocity 
 $\mathbf{u}_0(\mathbf{x})$ at the particle position $\mathbf{x}(t)$, and gravity, inertia as well as flow unsteadiness can be treated as a perturbations  \cite{Haller2008}. Under these conditions we obtain a reduced dynamical equation for inertial particles (see also Refs.  \citep{Robinson1956,Maxey1987jfm}):
\beq
 \dot{{\mathbf{x}}}(t) = {\mathbf{v}}(t) \simeq \mathbf{u}_0(\mathbf{x}) +   \epsilon  \mathbf{u}_1(\mathbf{x},t)  + \frac{St}{Fr}  \, \hat {\mathbf{g}}    - St \, (\mathbf{u}_0 . \nabla) \mathbf{u}_0  ,
\label{asymptcl1}
\eeq
plus quadratic combinations of $\epsilon$ and $St$. This equation will be used below.

We now consider a separatrix streamline (also denoted as dividing streamline)  of the steady flow $\mathbf u_0$ 
joining two saddle stagnation points $A$ and $B$,   with the flow from $A$ to $B$ (Fig. \ \ref{Dessin}). In the following, the separatrix will be simply denoted by $AB$.
The dynamics near  $AB$ can be investigated by using classical methods originally developed for separatrix maps \citep{zaslavsky1968,Kuznetsov1997,Chirikov1979}. We consider a solution $\mathbf{x}(t)$ of the complete perturbative equation (\ref{asymptcl1}), and determine the variations of the  unperturbed streamfunction $\psi_0[\mathbf{x}(t)]$ (defined by $\mathbf u_0 = \nabla \times (\psi_0 \, \hat {\mathbf z})$) along this perturbed trajectory.  Also, we will consider the "undisturbed streamfunction seen by the particle", that is the variation of  
$\psi_0$ at discrete  times $\tau_n$ and $\tau_{n+1}$, where $\tau_n$ and $\tau_{n+1}$ are  two consecutive times when the trajectory $\mathbf{x}(t)$ passes closest to the stagnation points $A$ and $B$ respectively:
\beq
\Delta \psi_0^p = \psi_0(\mathbf{x}(\tau_{n+1})) - \psi_0(\mathbf{x}(\tau_{n})) .
\eeq
Using Eq.\ (\ref{asymptcl1}) this quantity reads:
\beq
\Delta \psi_0^p = \int_{\tau_n}^{\tau_{n+1}} 
\nabla \psi_0  . \Big( \epsilon\mathbf{u}_1(\mathbf{x},t)  + \frac{St}{Fr}  \, \hat {\mathbf{g}}    - St \, (\mathbf{u}_0 . \nabla) \mathbf{u}_0
\Big) \, dt.
\eeq
The sign of $\Delta \psi_0^p$ contains direct information about the behavior of the particle near the heteroclinic cycle containing $AB$: a constant  sign will indicate that the particle penetrates  (or exits)  the cell once and for all, whereas oscillating signs correspond to a chaotic  motion: particles cross the separatrix $AB$ in a complex manner (see Chirikov \cite{Chirikov1979}, Guckenheimer \& Holmes  \cite{GH83},   Del Castillo-Negrete \cite{delcastillo1998}).  
 
  The variation of the undisturbed streamfunction  $ \Delta \psi_0^p$ is calculated as follows.
  We consider a cartesian coordinates system $(x,y)$ in the plane of the flow, with unit vector $\hat {\mathbf x}$ in the horizontal direction, and unit vector $\hat {\mathbf y}$ directed upward (that is, $\hat {\mathbf y}=- \hat {\mathbf g}$).
  Let $t_n \in ]\tau_n,\tau_{n+1}[$ be the time when $\mathbf{x}(t)$ passes nearest to some reference  point $C$ on  separatrix $AB$ (with $C$ not equal to $A$ nor $B$). By  
  writing that $\mathbf{x}(t) \simeq \mathbf{q}(t-t_n)$, where $\mathbf{q}(t)$ is an exact solution of the unperturbed dynamics on $AB$ with $\mathbf q(0)=C$, we are led to:
 \beq
\Delta\psi_0^p \simeq  \Delta \psi_0^f + \frac{St}{Fr} (\mathbf{AB} \times \hat {\mathbf g})\, . \, \hat {\mathbf z}
 - St \,  \langle | \mathbf u_0 | ^2   \rangle_{c},
\label{Deltapsi0}
\eeq
where $\mathbf{AB}$ is the displacement vector joining $A$ and $B$, and:
\begin{equation}
\Delta \psi_0^f = \epsilon\, \hat {\mathbf z} . \int_{-\infty}^{+\infty} \dot{\mathbf{q}} \times  \mathbf{u}_1(\mathbf{q}(t),t+t_n) \, dt,
\label{DeltaPsi0fExact}
\end{equation}
is the variation of streamfunction for fluid points, and
\begin{equation}
 \langle | \mathbf u_0 | ^2   \rangle_{c} = \hat {\mathbf z} . \int_{-\infty}^{+\infty} \dot{\mathbf{q}} \times \ddot {\mathbf{q}} \, dt
 \label{u0cdef1}
\end{equation}
is related to centrifugal effects and will be discussed below. In these calculations we have introduced the unit vector $\hat {\mathbf z} = \hat {\mathbf x} \times \hat {\mathbf y}$, which is perpendicular to the plane of the flow. Also,
$(\mathbf{AB} \times \hat {\mathbf g}) \, . \,\hat {\mathbf z} = -(\mathbf{AB} \times \hat {\mathbf y}) \, . \, \hat {\mathbf z} = -\mathbf{AB}  . \hat {\mathbf x}$, and will be denoted as $-x_{AB}$ in the following.

These results have been obtained by making use of the fact that, by definition of $\mathbf{q}$, we have $\dot{\mathbf{q}} = \mathbf{u}_0$ and $\ddot{\mathbf{q}} = (\mathbf{u}_0 . \nabla) \mathbf{u}_0$.  Also, the dynamics of $\mathbf{q}(t)$ is very slow near the stagnation points $A$ and $B$, so that integrals over $[\tau_n - t_n ,  \tau_{n+1} - t_n]$ have been replaced by integrals over $[-\infty,\infty]$ \citep{Chirikov1979,Kuznetsov1997}.  
 The coefficient $ \langle | \mathbf u_0 | ^2   \rangle_{c}$
is the effect of the curvature of  separatrix $AB$ on the dynamics of inertial particles. Indeed, the  local curvature  is  $R^{-1}= \hat {\mathbf z} . \dot{\mathbf{q}} \times \ddot {\mathbf{q}} / 
| \dot{\mathbf{q}} |^3$ so that $\langle | \mathbf u_0 | ^2   \rangle_{c}$ is exactly zero for straight separatrices, and is strictly non-zero for separatrices with a constant curvature. Inertial particles, experiencing a centrifugal force along curved separatrices, tend to drift towards the exterior of the recirculation cell, and this effect brings a constant contribution into $\Delta \psi_0^p$. The coefficient $ \langle | \mathbf u_0 | ^2   \rangle_{c}$ can be re-written as a curvilinear  integral along the separatrix:
\begin{equation}
 \langle | \mathbf u_0 | ^2   \rangle_{c} = \int_{-\infty}^{+\infty}  | \dot{\mathbf{q}} |^3 \frac{dt}{R(t)} = \int_A^B | \mathbf u_0 |^2 \, \frac{ds}{R(s)}
 \label{u0cdef2}
\end{equation}
where $s$ is the arc-length, and we made use of the fact that $| \dot{\mathbf{q}}  | dt = | d {\mathbf{q}}  | = ds$. The last integral is a curvature-weighted integral of the undisturbed squared fluid velocity, and this is why we have chosen to denote it by the bracket $\langle \cdot \rangle_{c}$.
Note that this integral is not always positive, as the curvature is a signed quantity: when  non-zero, $R^{-1}$ is positive for anti-clockwise streamlines, and negative otherwise.
Under these conditions Eq.\ (\ref{Deltapsi0}) leads to:
\begin{equation}
\Delta\psi_0^p \simeq  \Delta\psi_0^f  - \frac{St}{Fr} \, x_{AB}
 - St \, \langle | \mathbf u_0 | ^2   \rangle_{c} .
 \label{Deltapsi0bis}
\end{equation}
Eq.\ (\ref{Deltapsi0bis}) shows that flow unsteadiness, gravity and curvature bring additive contributions into $\Delta\psi_0^p$.  A positive   $\Delta\psi_0^p$ means that particles moving in the vicinity of  separatrix $AB$ will drift, during the time interval $[\tau_n,\tau_{n+1}]$, towards the left of the streamline (i.e. towards $\hat {\mathbf n}$, see Fig.\ \ref{Dessin}(i)). Negative contributions correspond to particles drifting towards its right-hand-side (i.e. towards $-\hat {\mathbf n}$). The flow unsteadiness
brings a varying sign. The contribution of gravity has the sign of $- x_{AB}$ and, since particles are heavier than the fluid here, it always corresponds to particles driven downward. The curvature-weighted integral of the squared velocity $\langle | \mathbf u_0 | ^2   \rangle_{c}$ is positive (negative) if $AB$ forms an anti-clockwise (clockwise) curve. In both cases it corresponds to a centrifugal effect.

Finally, let $\epsilon \, \alpha$ denote the peak value of the variation of streamfunction  of perturbed fluid points
trajectories along the
separatrix:
$$
\epsilon \, \alpha  = \max_{-\infty < t_n < \infty} | \Delta \psi_0^f (t_n) |
$$
A sufficient condition for  the sign of $\Delta \psi_0^p$ to remain constant  for all $n$ is:
 \begin{equation}
 St > St_{c} = \frac{\epsilon \, \alpha }{|x_{AB} / Fr + \langle | \mathbf u_0 |^2   \rangle_c| },
 \label{Stc}
 \end{equation}
where   $\alpha$   only depends on the structure of the  flow and manifests the effect of  unsteadiness. The curvature-weighted integral $ \langle | \mathbf u_0 |^2   \rangle_c$ also 
is independent of the particle characteristics. It
only depends on the geometry of the separatrix $AB$ and on the steady velocity distribution along this streamline.

All variables in the results above have been set non-dimensional by using some flow length-scale $L$ and velocity $U$.
The dimensional counterpart of Eq.\ (\ref{Stc}) is:
\begin{equation}
 \tau_p > \tau_{pc} = \frac{\delta}{| g \, X_{AB}  + \langle | \mathbf U_0 |^2   \rangle_c | },
 \label{taupc}
 \end{equation}
 where $\tau_p = St \times L/U$ is the response time of the inertial particle~; $X_{AB}$ is the dimensional $x$-coordinate of vector $\mathbf{AB}$~; $\mathbf U_0$ is the dimensional  undisturbed fluid velocity vector~;  $\delta = \epsilon\,\alpha \times U L$ 
 has the dimension of a diffusion coefficient
 ($m^2/s$). It is the flux of fluid points within the stochastic zone near  separatrix $AB$, due to the flow unsteadiness (see for example Beigie {\it et al.}  \citep{Beigie1991,Beigie1994}, Solomon {\it et al.}  \cite{Solomon2001}). Note that this quantity is not related to inertial particles and only depends on the structure and on the unsteadiness of the fluid flow.

The criterion in Eq.\ (\ref{taupc})  can also be written
as $D > D_c$, where $D_c$ is the critical diameter of the particle: 
\begin{equation}
D_c= \left( 18 \frac{\mu}{\rho_p} \tau_{pc} \right)^{1/2}.
\label{Dc}
\end{equation}
This expression will be used in the next sections, and compared to numerical simulation results.

\begin{figure}
{\includegraphics[width = .5\textwidth]{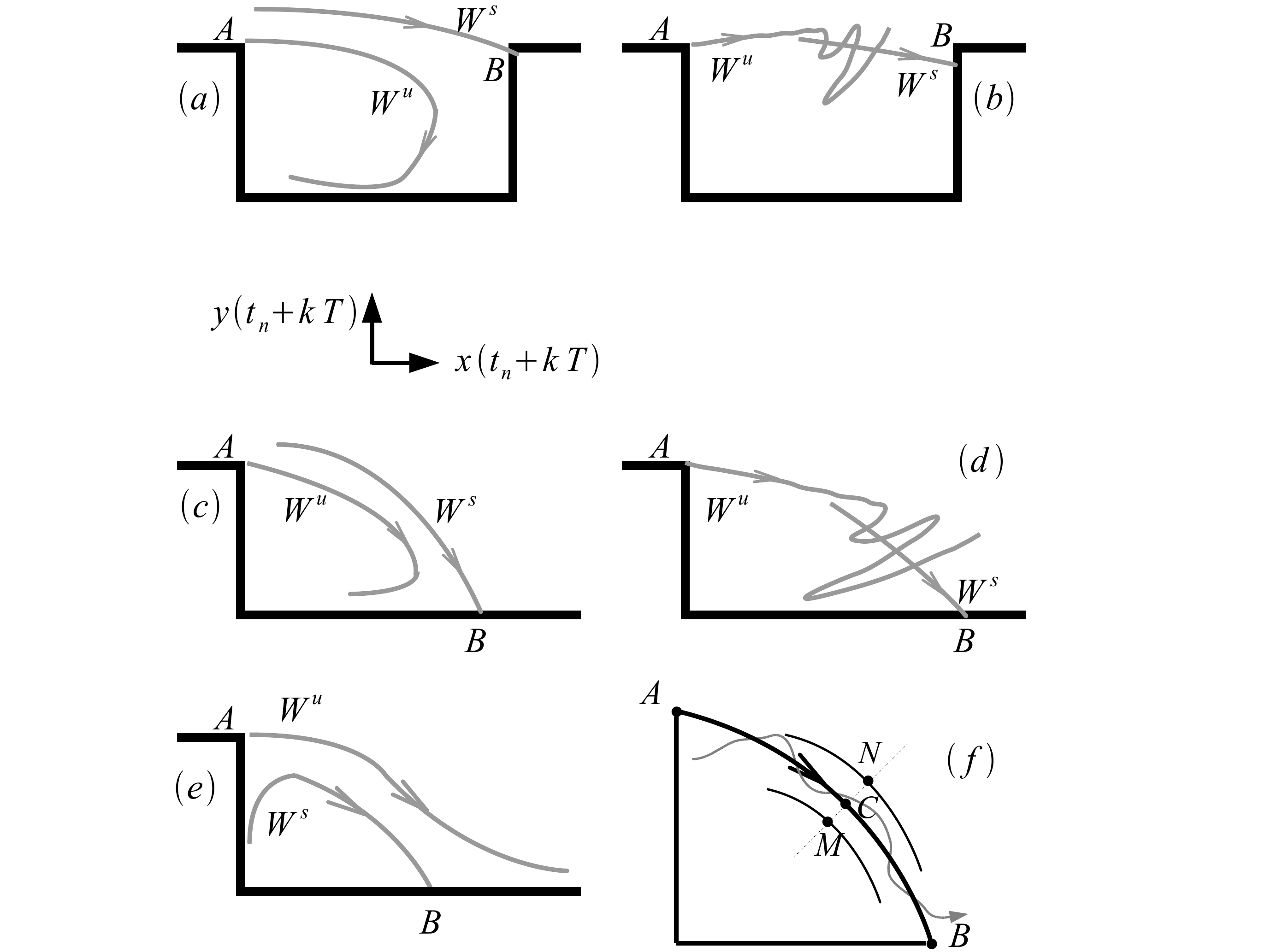}} 
\caption{\it Sketch of the invariant manifolds of the dynamics of particles in the cavity flow $(a)-(b)$ and in the backward facing step flow $(c)-(d)$, when particles are assumed to slip on the walls. In the steady case, or in the unsteady case with $St > St_c$ (that is $D>D_c$), manifolds do not intersect and take the form $(a)$ and $(c)$ if gravity dominates inertial curvature effects: the motion is regular and particles entering the cell cannot exit. In the unsteady case with $St < St_c$ (that is $D<D_c$), manifolds take the form $(b)$ and $(d)$: the motion is chaotic, some particles can enter and exit the cell in a complex manner. Case $(e)$ corresponds to a regular motion when centrifugal effect dominates. Panel $(f)$ is a sketch of a trajectory wandering in the stochastic zone near the separatrix.} 
\label{WuWs}
\end{figure}

{\it Structure of invariant manifolds.} To understand further the physical meaning of these analytical results, consider the structure of the invariant manifolds of particle's dynamics (sketched in Fig. \ref{WuWs}). Assuming that particles are allowed to slip on the wall, separatrix $AB$ and walls form a heteroclinic cycle. We consider
Poincar\'e sections of the particle dynamics: $\mathbf{x}(t_n+k T)$, where $k$ is an integer, $T$ is the period of some external perturbation and $t_n$ is any arbitrary time where the Poincar\'e section starts. Because the leading-order dynamics in (\ref{asymptcl1}) has hyperbolic saddle-points $A$ and $B$, the Poincar\'e map of the perturbed dynamics will also display hyperbolic points $A$ and $B$ in the vicinity of the undisturbed ones, the eigenvalues of which have the same signs than the unperturbed ones provided the perturbation is small enough. Therefore, an unstable invariant manifold $W^u$ emerges from $A$ and a stable manifold converges towards $B$. The two manifolds, which coincided with the separatrix streamline in the steady case, no longer coincide in the unsteady dynamics in general. The existence of intersection points between both manifolds is related to the existence of zeros in $\Delta\psi_0^p(t_n)$ (see for example Ref.  \cite{GH83}). If $\Delta\psi_0^p(t_n)$ is strictly negative for all $t_n$, then  manifolds do not intersect each other and  take the form  sketched in Fig.\ \ref{WuWs}$(a)$ and $(c)$ [in the case of the cavity and of the backward-facing step with $AB$ towards positive $x$]. Any low-$St$ particle, injected sufficiently close to $A$, will be driven towards the interior of the cell and never exit. 
This happens in the steady case, or in the unsteady case with $St > St_c$.
In contrast, if $\Delta\psi_0^p(t_n)$ has simple zeros when $t_n$ varies, then the manifolds will have an infinite number of intersection points (Fig.\ \ref{WuWs}$(b)$ and $(d)$): particles injected near $A$ can have a chaotic dynamics near the  heteroclinic loop, and enter and exit the cell in an unpredictable manner. This kind of behaviour happens in the unsteady case when $St < St_c$.

Note that the regular behaviour can also happen when centrifugal effects dominate gravity and unsteadiness. Indeed, if $AB$ is clockwise ($\langle | \mathbf u_0 |^2 \rangle < 0$) and $-  \langle | \mathbf u_0 |^2 \rangle $ dominates both the unsteady term $\Delta\psi_0^f$ and the gravity term $x_{AB}/Fr$, then  
$\Delta \psi_0^p(t_n)$ is always positive:   particles will be centrifuged away irrespective of their Stokes numbers, provided it is small enough. In this regime, sketched in Fig. \ref{WuWs}$(e)$,   any particle injected close enough to the heteroclinic cycle will be centrifuged away.

\section{Application to an analytical field and validation of the numerical approach}
\label{illustrationAnalyt}

The method, and its applicability to numerical velocity fields, is first tested by means of an elementary analytical velocity field:
$$
\psi(x,y,t) = \psi_0(x- \epsilon \sin \omega t,y ) 
$$
\begin{equation}
\simeq \psi_0(x,y) + v_0(x,y) \epsilon \sin \omega t + O(\epsilon^2)
\label{psiexact}
\end{equation}
where the steady (nondimensional) streamfunction $\psi_0$ is taken to be:
\begin{equation}
\psi_0(x,y)  = y (x^2+y^2-1),
\label{psi0}
\end{equation}
and $v_0(x,y) = -\partial \psi_0/\partial x$ is the corresponding vertical velocity. This steady flow is sketched in Fig. \ref{Dessin}$(ii)$: it consists in a pair of half-circular cells with a unit radius. The perturbed streamfunction $\psi$ corresponds to a periodic rigid-body translation of the flow (\ref{psi0}) in the horizontal direction.

We first consider the upper dividing streamline, that is the upper half circle $AB$ in the  half plane $y>0$. Gravity is along $-y$: it pushes particles within the cell,
and is opposed to  centrifugal effects. By inserting the analytical expression of $u_0(x,y)$ into Eq.\ (\ref{u0cdef2}), and setting $x= \sin \phi$ and $y=\cos \phi$,  the curvature-weighted integral is readily calculated. We obtain
$
\langle |\mathbf u_0 |^2   \rangle_c 
= -2 \pi$,
the negative sign being due to the fact that the upper dividing streamline is clockwise.
Moreover, for this simple flow, a solution $\mathbf{q} = (q_x(t),q_y(t))$ of the unperturbed dynamics on $AB$ can be readily obtained, and we get
\begin{equation}
q_x(t) = \tanh 2 t  \quad \mbox{and} \quad q_y(t) = \frac{1}{\cosh 2 t}.
\end{equation}
By inserting these functions into (\ref{DeltaPsi0fExact}) we obtain an exact expression for $\Delta\psi_0^f$, and:
\begin{equation}
\Delta\psi_0^p(t_n) \simeq   \frac{\pi}{2} \frac{\omega^2 \, \epsilon}{\cosh (\pi \omega/4)} \sin \omega t_n  + 2 St( \pi - \frac{1}{Fr}), 
 \label{Deltapsi0CellUpper}
\end{equation}
where we have used $x_{AB}=2$.
Therefore, in the steady case ($\epsilon=0$),   $\Delta\psi_0^p(t_n)$ is of the sign of $(\pi - \frac{1}{Fr})$ for all $t_n$: this means that, if $Fr<1/\pi$ gravity drives inertial particles inside cell irrespective of the (small) Stokes number $St$. In contrast, if
$Fr > 1/\pi$ centrifugal effects dominate and particles slip on the arc $AB$ and slowly drift away from it: any low-St particle coming from the upstream region "bounces" on the cell. This last situation corresponds to the case where the unstable manifold $W^u$ emerging from $A$ is towards the exterior of the cell and never intersects $W^s$ [as sketched in Fig.\ \ref{WuWs})(e) for the step]. Note that a peculiar case emerges if $Fr=1/\pi$: here gravity and centrifugal effect balance each other and $\Delta\psi_0^p(t_n)$ is an oscillating function, whatever the (small) Stokes number is. 

For the lower dividing streamline $AB$ of the circular cell, i.e. the half-circle $AB$ in the plane $y < 0$, we have $
\langle |\mathbf u_0 |^2   \rangle_c 
= +2 \pi$, and we get
\begin{equation}
\Delta\psi_0^p(t_n) \simeq  - \frac{\pi}{2} \frac{\omega^2 \, \epsilon}{\cosh (\pi \omega/4)} \sin \omega t_n  - 2 St( \pi + \frac{1}{Fr}). 
 \label{Deltapsi0CellLower}
\end{equation}
Hence, the critical Froude number does not exist here, and both  gravity and centrifugal effects push the particles out of the cell, downward. 

\begin{figure}
{\includegraphics[width = .5\textwidth]{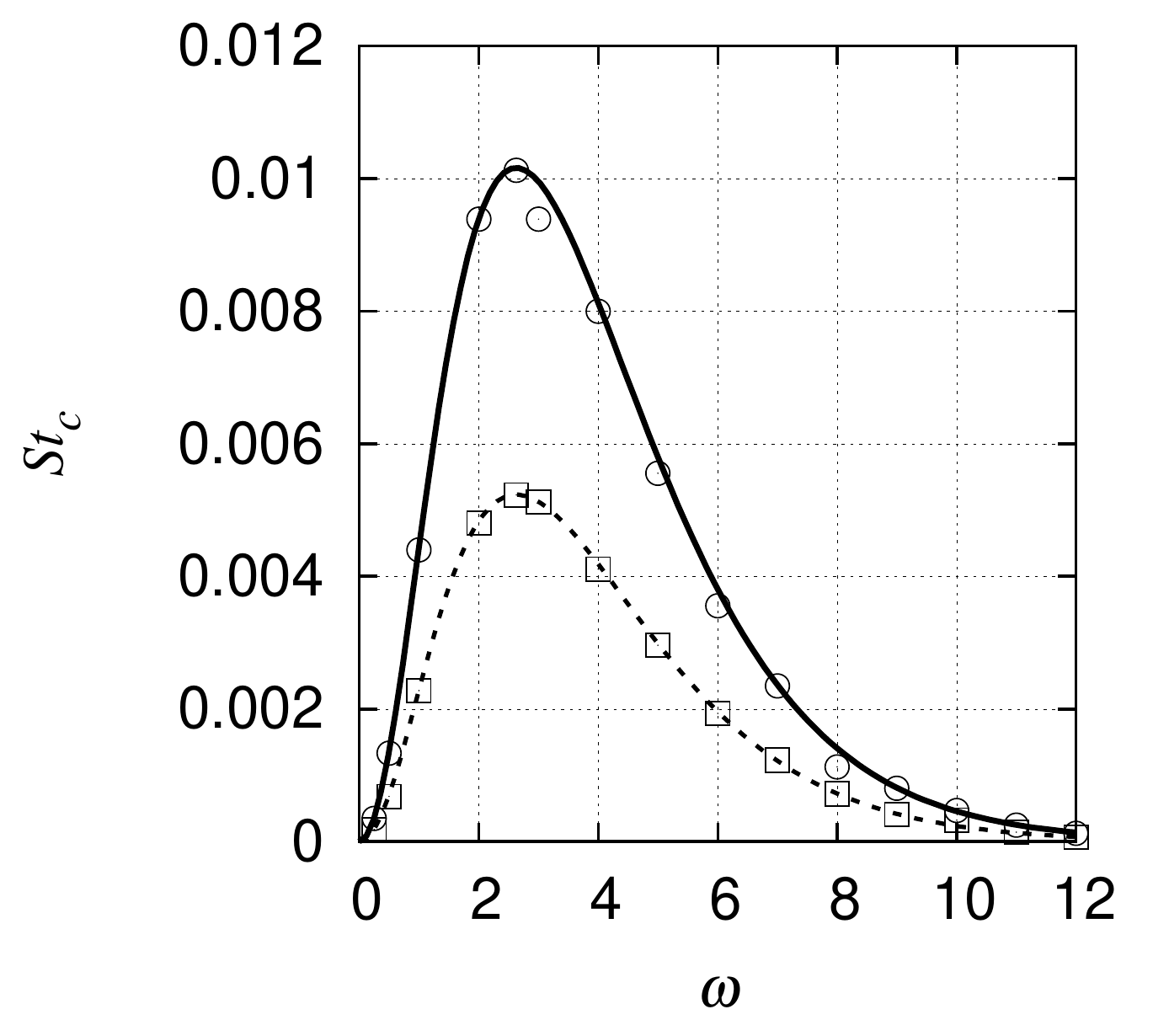}} 
\caption{\it Continuous lines : critical Stokes numbers $St_c^+$ (solid line) and $St_c^-$ (dashed line), for the circular separatrices of the analytical cell flow (\ref{psiexact}), obtained from the theoretical expressions (\ref{Stccercle}). Symbols : the critical Stokes numbers obtained from numerical simulations of a large number of particle trajectories in the same flow  discretized on a mesh. Numerical values of the parameters in these simulations: $\rho_p/\rho_f = 1000$,  $Fr=0.1$, $\epsilon=0.05$.  } 
\label{StcAnalytNumer}
\end{figure}

In the unsteady case ($\epsilon \not =0$), formulas (\ref{Deltapsi0CellUpper}) and (\ref{Deltapsi0CellLower}) imply that  
 the undisturbed streamfunction seen by the particle oscillates with $t_n$ if:
\begin{equation}
St < St_c^\pm(\omega) = \frac{\pi}{|\pi \pm 1/Fr|}\frac{\omega^2 \, \epsilon}{4 \cosh (\pi \omega/4)},
\label{Stccercle}
\end{equation}
the $\pm$ sign corresponding to the upper (+) and lower (-) separatrix respectively.
The curves $St_c^+(\omega)$ and $St_c^-(\omega)$ have been plotted in Fig. \ref{StcAnalytNumer} when $Fr=0.1$ ($< 1/\pi$) and $\epsilon=0.05$: they have a maximum when $\omega \simeq 2.63$, and decay rapidly when $\omega$ increases above this value. These curves will be compared to numerical results below.  Finally, the theory can be applied to the horizontal separatrix of the cellular flow (with $A$ playing the role of $B$ and vice versa), and we readily obtain $\Delta\psi_0^p(t_n) = 2 St/Fr$ for all $n$. This means that, for such a horizontal perturbation, only gravity matters and drives the particles towards the lower half-cell during their motion from $B$ to $A$.

\begin{figure}
{\includegraphics[width = .5\textwidth]{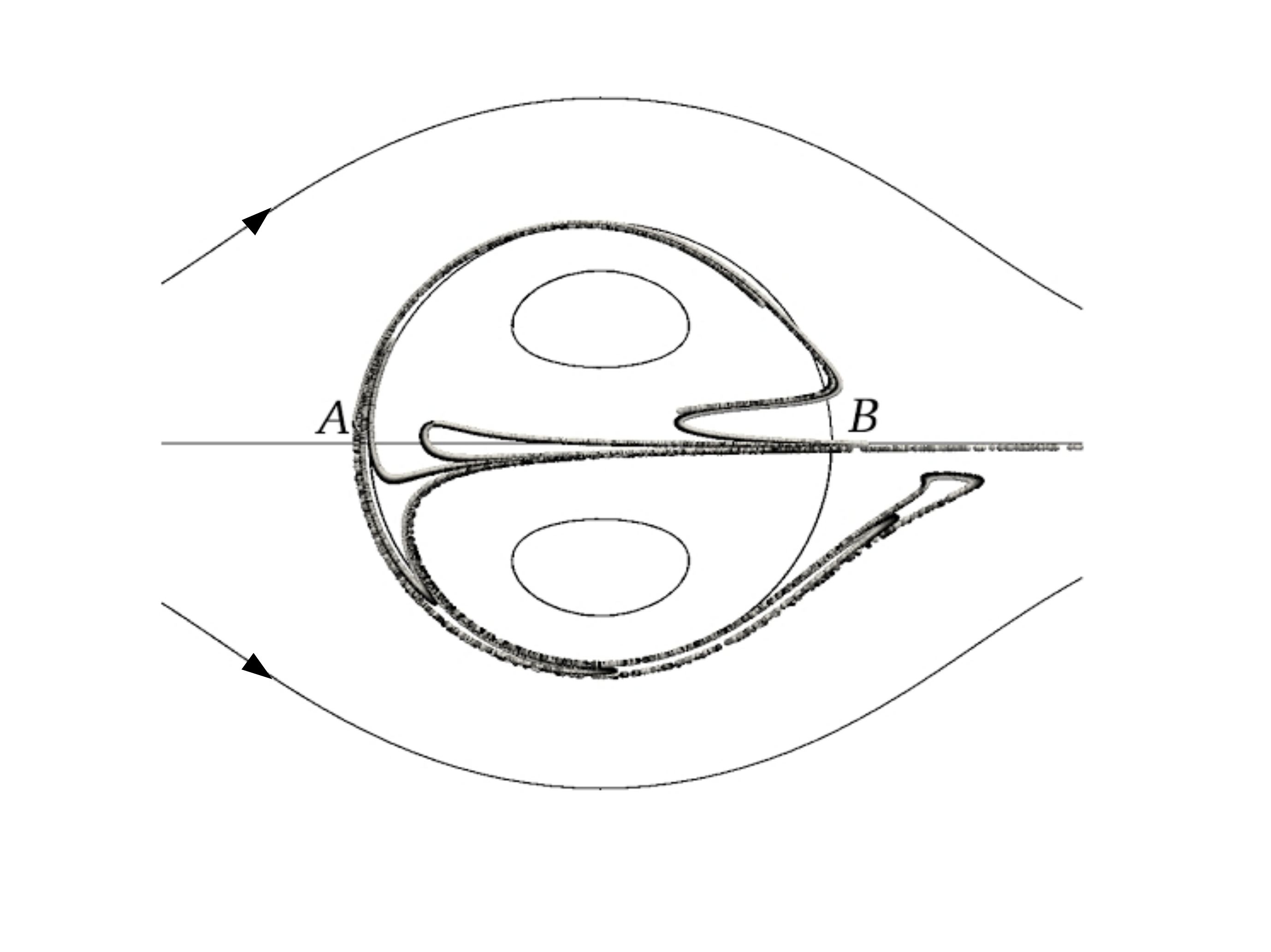}} 
 \vspace*{-.5cm}
\caption{\it Test-case : typical clouds of particles transported in the vicinity of the unsteady cell flow (\ref{psiexact}), and initially injected near point~$A$, with $y>0$. Six particle diameters are used, corresponding to six different values of the Stokes number.  
All Stokes numbers are super critical for the lower separatrix (that is: $St_c^- < St$). Hence, particles who reach the lower circular separatrix $AB$ are too inertial to re-enter the cell. Numerical values of the parameters in these simulations: $\rho_p/\rho_f = 1000$,  $Fr=0.1$, $\epsilon=0.05$.  
} 
\label{CelluleNuage}
\end{figure}

Because we want to determine the critical Stokes number of inertial particles advected by {\it numerical } velocity fields,  we have performed numerical computations of particle trajectories using a discretization of the
analytical velocity field (\ref{psiexact}). Typical clouds with various diameters
 are shown in Fig.\ \ref{CelluleNuage}. 
We then have checked whether or not particles have an irregular dynamics around the dividing streamlines, for various $\omega$'s between 0 and 12, and recorded the corresponding critical Stokes number below which this happened. This $St_c$ has then been compared to the theoretical expression (\ref{Stccercle}) in  Fig. \ref{StcAnalytNumer}. We observe that the agreement is good, so that the asymptotic analysis can be applied to numerical velocity fields also.

The application of the asymptotic theory to numerical fields requires that the positions of points $A$ and $B$, and the curvature-weighted integral $\langle |\mathbf u_0 |^2   \rangle_c$, be determined from the discrete velocity field. In the case of the cellular flow investigated in this section,
 separatrix $AB$ has been determined approximately by calculating a fluid point trajectory in the steady velocity field ($\epsilon=0$), injected very close to point $A$.
The curvature-weighted integral $\langle | \mathbf u_0 |^2   \rangle_c$ is then computed. For our finest mesh we obtain
$
\langle |\mathbf u_0 |^2   \rangle_c \simeq -6.28,
$
which agrees reasonably with the expected value $-2\pi$. 

Note however that the computation of $\langle | \mathbf u_0 |^2   \rangle_c$ done in this section could also be performed from any velocity field obtained by solving the Navier-Stokes equations with a numerical solver. Then, the  behaviour of particles near the separatrix can be predicted by simply applying formula (\ref{Stc}), or its dimensional counterpart (\ref{taupc}), without doing any inertial particle tracking. This is done in the next section, in the case of a cavity flow  and of a backward-facing step.

\section{Application to numerical velocity fields}
   \label{Numerical}
   
To illustrate further these results and show their applicability to numerical solutions of the Navier-Stokes equations, we consider the classical examples sketched in Figs.\ \ref{Dessin}$(iii)$ and \ref{Dessin}$(iv)$. We claim that the behaviour of particles can be analysed by calculating the various terms of Eq.\ (\ref{Deltapsi0bis}), re-written here in dimensional form (keeping the same symbols for streamfunctions):
\begin{equation}
\Delta\psi_0^p \simeq  \Delta\psi_0^f  - g \, \tau_p \, X_{AB}
 - \tau_p \, \langle | \mathbf U_0 | ^2   \rangle_{c}.
 \label{Deltapsi0ter}
\end{equation}
{\it Numerical method.} For both the cavity and the backward facing step, numerical simulations using a finite-volume algorithm  have been performed on a structured mesh using Gauss integration and linear interpolations. The prediction-correction PISO algorithm has been used  \cite{Issa1986}. Time stepping is done by means of a   Crank-Nicolson scheme, and a CFL number (Courant-Friedrichs-Lewy) below 0.1 is used in all cases.  
Particles are calculated by means of linear interpolation coupled to a Euler time-stepping method. 
They are allowed to slip on the walls, so that the separatrix streamline and the wall form a heteroclinic loop. 
For each geometry, two runs have been performed. First, a steady boundary condition has been used, together with a moderate Reynolds number, to obtain a steady velocity field with streamlines of the form of those sketched  in Fig.\ \ref{Dessin}. Then, the separatrix and the curvature-weighted integral  $\langle | \mathbf U_0 |^2   \rangle_c$ have been computed, and injected into formula (\ref{taupc}).  This equation also requires to determine the  flux  $\delta$. This quantity, which is only flow dependent, can be estimated after a run involving fluid points only, since the variation of streamfunction is also the flow rate between the two corresponding streamlines. In the vicinity of any point $C$ of the separatrix we have:
$$
\delta = \max_{t_n} |\Delta\psi_0^f(t_n)| \sim |\psi_0(M)-\psi_0(N)| 
$$
where $M$ and $N$ denote the intersection between the border of the stochastic zone and the line normal to separatrix $AB$ at $C$ (see Fig.\ \ref{WuWs} (f)). A first order expansion of $\psi_0(M)-\psi_0(N)$ leads to
\begin{equation}
\delta  \sim    d \, | \mathbf{U}_0(C) |,
\label{Estimedelta}
\end{equation}
  where $d = |MN|$ is the thickness of the stochastic layer at $C$. By measuring the thickness $d$ from fluid point simulations, we estimate the flux of fluid points $\delta$ from formula (\ref{Estimedelta}), and the critical   response time $\tau_{pc}$, or diameter $D_c$ from Eqs.\ (\ref{taupc}) and (\ref{Dc}). We have checked that the exact position of $C$ on the separatrix does not affect the resulting $\delta$, as expected.
These predictions have then been compared to those obtained by tracking a large number of particles injected near $A$, and by checking whether separatrix crossing occurred.

\begin{figure}
{\includegraphics[width = .5\textwidth]{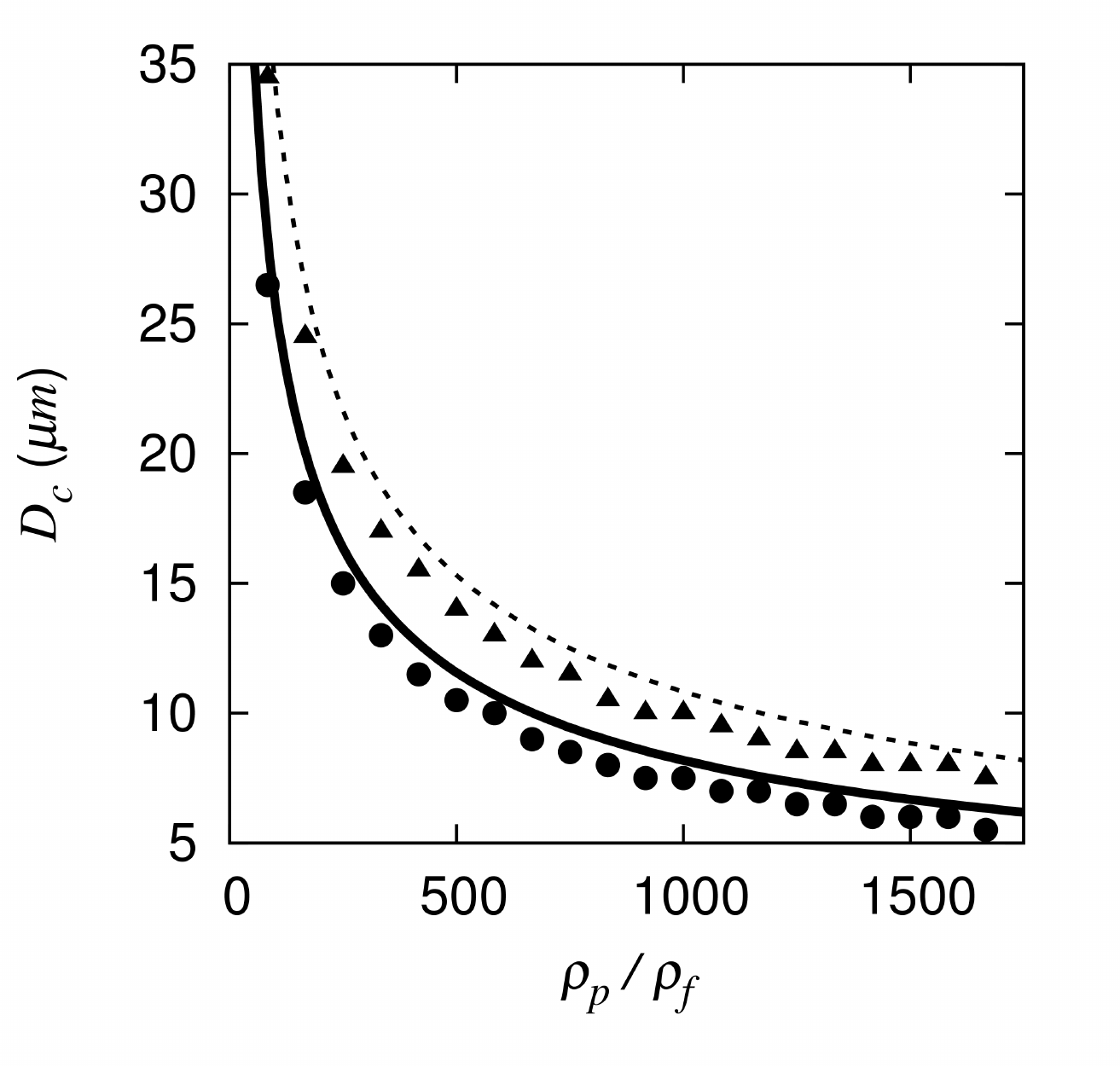}} 
\caption{\it Critical diameter obtained from the asymptotic theory (solid line : cavity, dashed line : backward-facing step). Symbols indicate the diameters obtained by tracking a large number of inertial particles (circles : cavity, triangles : backward-facing step). Numerical values of the parameters:  $\omega = 26.2\, rad/s$, $\epsilon=0.05$.   } 
\label{Dcfig}
\end{figure}
 
\begin{figure}
 {\includegraphics[width = .5\textwidth]{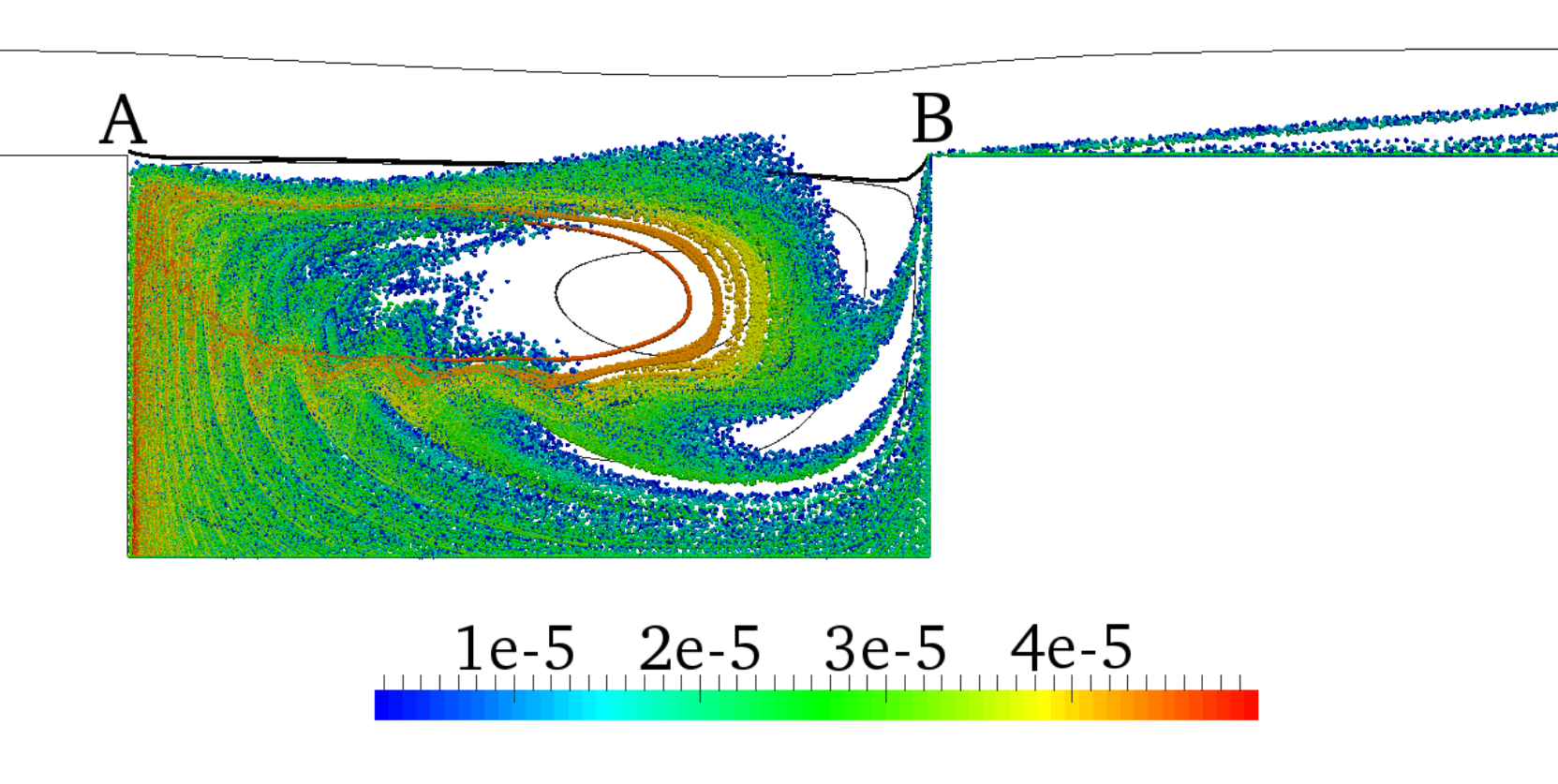}} 
\caption{\it Typical clouds of particles moving in the vicinity of the cavity flow. 
The thick line is the dividing streamline $AB$ of the steady flow.
Colours indicate particles' diameter $D$, in $m$. All particles are injected near $A$, and their density is $\rho_p=100 \, kg/m^3$. 
Those for which $D > D_c = 26.5$ microns have a regular dynamics  and cannot exit.  Other parameters: $U_\infty^0 = 0.13\, m/s$, cavity size $1\, cm \times 2\, cm$, $\omega = 26.2\, rad/s$, $\epsilon=0.05$. } 
\label{snapcav}
\end{figure}

{\it Cavity flow.} We first consider the flow above a two-dimensional open cavity sketched in Fig. \ref{Dessin}$(iii)$. 
This flow  has been widely investigated, in various contexts  \citep{liu2002large,Caton2003,lusseyran2008dynamical}.
 The upstream flow has an imposed parabolic velocity profile with a cross-sectional average velocity $U_\infty^0 = 0.13\, m/s$. The depth and width of the cavity are $L = 0.01 \, m$ and 
$W=0.02 \, m$ respectively. The corresponding
     Reynolds number, based on the viscosity of air at 20$^o$C, is $Re = U_\infty^0 L/\nu \approx 90$, which corresponds to a moderate value. The flow is steady and characterized by a separatrix streamline $AB$ separating open streamlines (above the cavity) from closed streamlines (in the cavity).  
   We have chosen to introduce some unsteadiness by imposing a pulsatile velocity upstream of the cavity:
\begin{equation}
U_\infty(t) = U_\infty^0 (1 + \epsilon \sin \omega t)
\label{UinfInstat}
\end{equation}
with $\omega = 26.2\, rad/s$,  
which corresponds to a period of the order of the convective time over the cavity,   and $\epsilon = 0.05$. 
 The question of interest here is to determine under which conditions  aerosols released outside (near point $A$)
   can be captured within the cavity, and remain there either permanently or temporarily. By computing a lagrangian fluid point trajectory near $AB$ we get  $\langle \, |\mathbf U_0 |^2  \, \rangle_c \simeq 2.5\,10^{-4} \, m^2/s^2 \ll g X_{AB} = g \, W \simeq 0.2\, m^2/s^2$, so that gravity dominates and Eq.\ (\ref{Deltapsi0ter}) reads
   \begin{equation}
\Delta\psi_0^p \simeq  \Delta\psi_0^f  - g \, \tau_p \, X_{AB}.
\end{equation}
We therefore conclude that, in the steady case ($\Delta\psi_0^f(t_n)=0$ for all $n$), the structure of the manifolds is similar to the one sketched in Fig.\ \ref{WuWs}$(a)$, and particles injected sufficiently close to $A$ drop regularly in the cell and cannot exit. This behaviour persists in the unsteady case provided $\tau_p > \tau_{pc}$ given by
\begin{equation}
  \tau_{pc} \simeq \frac{\delta}{ g \, W   }.
  \label{taupc_cavite}
\end{equation}
In contrast, particles such that $\tau_p < \tau_{pc}$, injected close enough to the separatrix, will have a chaotic dynamics and wander in and out.
 The flux of fluid points along the stochastic zone $\delta$ has been obtained by measuring the thickness $d$ of the stochastic zone around the separatrix $AB$, and using formula (\ref{Estimedelta}).  
 We obtain 
 $\delta \simeq 5.0\, 10^{-5} \, m^2/s$. This leads to a critical response time $\tau_{pc} \simeq 2.5 \, 10^{-4} \, s$. To check these predictions we have computed inertial particle trajectories with varying density ratios $\rho_p/\rho_f$. We then have controlled whether particles can exit the cavity or not. The critical diameter below which this happens has   been plotted versus the density ratio (Fig.\ \ref{Dcfig} (circles)). The analytical $D_c$ (obtained from Eqs.\ (\ref{taupc_cavite}) and (\ref{Dc})) has been plotted on the same graph (solid line), we observe that the agreement is satisfactory. The systematic overestimation of the theoretical diameter, compared to the numerical one, might be due to the numerical determination of the thickness of the stochastic layer $d$. Indeed, by measuring $d$ from fluid points simulations, we  slightly overestimate the actual thickness. Therefore, the theoretical critical diameter is overestimated too.
Snapshots of particle clouds with diameters $D \in [1,50]$ microns are shown in Fig.\ \ref{snapcav} in the case where $\rho_p = 100\, kg/m^3$. All particles are injected near $A$. The most inertial particles (in red), for which $D > D_c = 26.5$ microns, drop in the cavity and remain there permanently.
  
\begin{figure}
{\includegraphics[width = .5\textwidth]{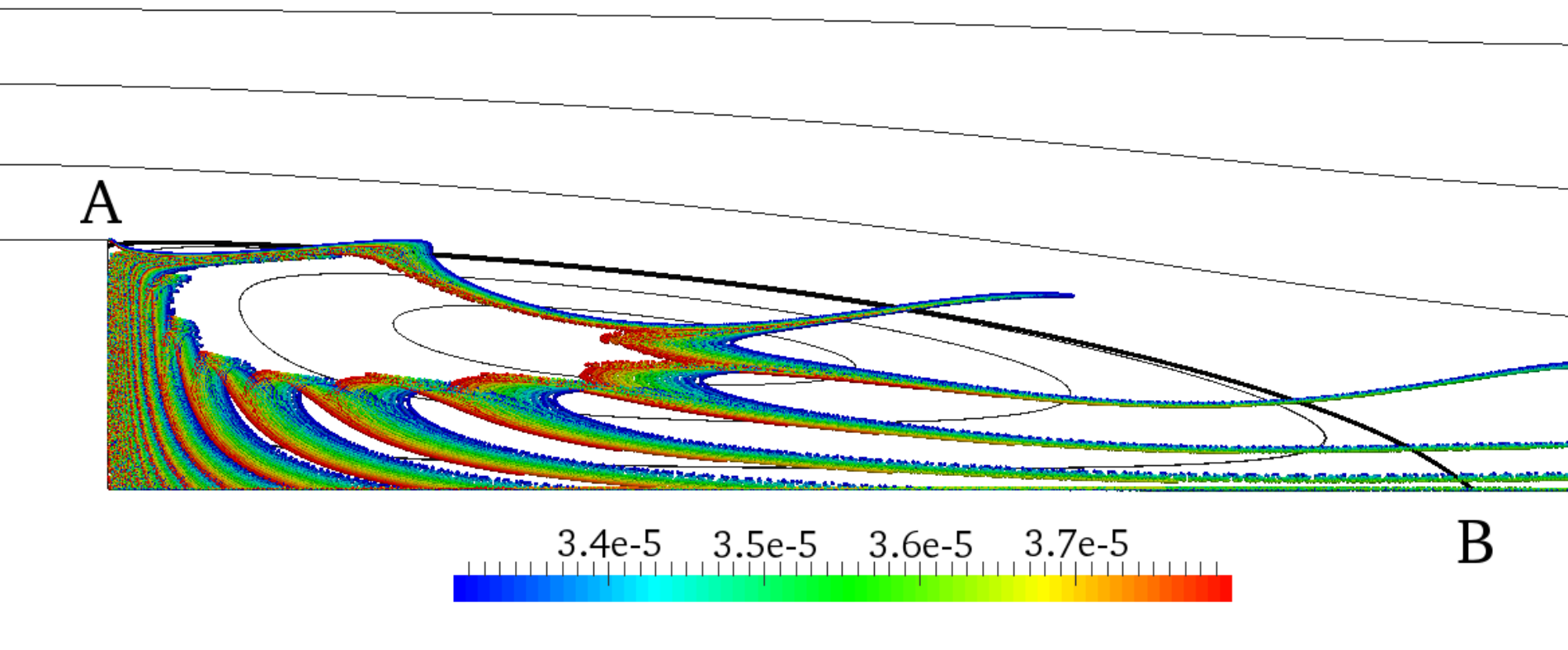}} 
\caption{\it Typical clouds of particles moving in the vicinity of the backward facing step. The thick line is the dividing streamline $AB$ of the steady flow.
 Colours indicate particles' diameter $D$, in $m$.  All particles are injected near $A$, and their density is $\rho_p=100 \, kg/m^3$. 
Those for which $D > D_c = 34.5$ microns (red particles) drop in the cell and are trapped permanently. They have a regular dynamics. Those for which  $D < D_c$ (blue) have a chaotic dynamics near the heteroclinic cycle and can exit in a complex manner. Other parameters: $U_\infty^0 = 0.17\, m/s$, vertical step size $L = 0.01 \, m$, $\omega = 25.2\, rad/s$, $\epsilon=0.05$.   } 
\label{snapBFS}
\end{figure}

{\it Backward-facing step.} A similar analysis has been done with the backward-facing step flow sketched in Fig.\ \ref{Dessin}$(iv)$.  This flow has been widely investigated numerically or experimentally
 \citep{yu2004numerical,armaly1983experimental,Ruck1988,kostas2002particle}.
 It is an interesting  configuration where centrifugal and gravity effects act in opposite directions if the step is downward. 
 Indeed, we have $g X_{AB} > 0$ and,
  because the dividing streamline is clockwise,   
  $\langle | \mathbf U_0 |^2   \rangle_c  < 0$.  In our simulations, the upstream flow is horizontal with a cross-sectional average velocity $U_\infty^0 = 0.17\, m/s$. The vertical size of the step is $L = 0.01 \, m$ and the Reynolds number is $Re = U_\infty^0 L/\nu \approx 115$. Here also, for such a moderate Reynolds number, the flow is steady. Also, $X_{AB} = 0.057 \, m$.  The gravity term appearing in Eq.\ (\ref{Deltapsi0ter}) is $g X_{AB} \simeq 0.56 \, m^2/s^2$ and the curvature term is $\langle \, |\mathbf U_0 |^2  \, \rangle_c \simeq  -0.00145 \, m^2/s^2 $. Gravity therefore dominates the dynamics of  particles near the separatrix: in the steady case the structure of the manifolds is similar to the one sketched in Fig.\ \ref{WuWs}$(c)$, and particles injected very close to $A$ drop regularly in the cell and cannot exit. This behaviour persists in the unsteady case provided $\tau_p > \tau_{pc}$ given by Eq.\ (\ref{taupc}).  In these simulations also unsteadiness is forced by imposing a pulsatile inlet velocity (Eq.\ (\ref{UinfInstat})).   
  The corresponding critical diameter is plotted in Fig.\ \ref{Dcfig} (triangles). The flux $\delta$ has been obtained by computing fluid points trajectories near the separatrix, and measuring the thickness of the stochastic zone. Then, by using formula (\ref{Estimedelta}) we get $\delta \simeq 2.5\, 10^{-4} \, m^2/s$.
  The corresponding critical diameter   is shown on the same graph (dashed line). Fig.\ \ref{snapBFS} shows particle clouds with various diameters between 32 and 38 microns.

  \vskip1cm
  
\section{Conclusion }
\label{concl}

The analysis presented in this study faithfully predicts the behaviour of low-Reynolds number heavy particles in the vicinity of the boundary of recirculation cells. It concerns the limit of vanishing Stokes numbers, that is particles with a finite but small inertia. For such particles, separatrix crossing depends on the competition between three mechanisms, namely flow unsteadiness, streamline curvature,  and gravity. The former effect always contributes to separatrix crossing and chaotic dynamics (if the separatrix is part of a homoclinic or heteroclinic cycle). The other effects can either contribute to separatrix crossing or be opposed to it, according to the geometry of the unperturbed flow. The criterion which has been derived in the present study enables to predict the various situations. It has been first confronted to numerical results involving an analytical flow discretized on a grid. Then, the criterion has been applied to numerical solutions of the Navier-Stokes equations, and we obtained satisfactory results.  

Note that the method can be applied to a wide variety of two-dimensional unsteady flows, even if the unsteadiness is more complex than the periodic time-dependence considered here. In particular, the method should work also for multimodal perturbations, or even turbulent flows, provided the turbulence intensity is weak (i.e. the typical scale $u'$ of velocity fluctuations is much smaller than the scale $U$ of the average velocity). The only contribution of the flow unsteadiness to separatrix crossing is contained in the flux $\delta$, which can be measured accurately from simulations of fluid point trajectories and Poincar\'e sections near the separatrix. 

The generalization of the asymptotic analysis to particles that are not much heavier than the fluid is the next step of this work. In this case additional contributions of the hydrodynamic force must be taken into account: pressure gradient of the undisturbed flow, added mass, history force, drag and lift due to the inertia of the displaced fluid. Even though the exact expression of the last two forces is still under debate, they are known to affect the complex dynamics of the particles (see for example Guseva {\it et al.}  \cite{Guseva2013} for a detailed analysis of the effect of the history force). The generalization of the asymptotic analysis to such particles would be useful for the prediction of tiny objects in water flows.

\vskip.1cm
{\bf Acknowledgement.} JRA's research was partially supported by NSF Grant PHY-1001198 while
visiting the department of Physics and Astronomy at Northwestern University. This work benefited from fruitful discussions with Prof. Adilson Motter.


\end{document}